# Coupling Between Magnetic and Transport Properties in Magnetic Layered Material Mn$_{2-x}$Zn$_x$Sb

Md Rafique Un Nabi[1,2], Rabindra Basnet[1], Krishna Pandey[3], Santosh Karki Chhetri[1], Dinesh Upreti[1], Gokul Acharya[1], Fei Wang[4], Arash Fereidouni[1,2], Hugh O. H. Churchill[1,2,3], Yingdong Guan[5], Zhiqiang Mao[5], Jin Hu[1,2,3*]

[1]Department of Physics, University of Arkansas, Fayetteville, AR 72701, USA

[2]MonArk NSF Quantum Foundry, University of Arkansas, Fayetteville, Arkansas 72701, USA

[3]Materials Science and Engineering Program, Institute for Nanoscience and Engineering, University of Arkansas, Fayetteville, AR 72701, USA

[4]Department of Chemistry and Biochemistry, Missouri State University, Springfield, MO 65897, USA

[5]Department of Physics, Pennsylvania State University, University Park, PA 16802, USA

## Abstract

We synthesized single crystals for Mn$_{2-x}$Zn$_x$Sb ($0 \leqslant x \leqslant 1$) and studied their magnetic and electronic transport properties. This material system displays rich magnetic phase tunable with temperature and Zn composition. In addition, two groups of distinct magnetic and electronic properties, separated by a critical Zn composition of $x = 0.6$, are discovered. The Zn-less samples are metallic and characterized by a resistivity jump at the magnetic ordering temperature, while the Zn-rich samples lose metallicity and show a metal-to-insulator transition-like feature tunable by magnetic field. Our findings establish $Mn_{2-x}Zn_xSb$ as a promising material platform that offers opportunities to study how the coupling of spin, charge, and lattice degrees of freedom governs interesting transport properties in 2D magnets, which is currently a topic of broad interest.

* jinhu@uark.edu

# 1. Introduction

Electronic transport properties are correlated with the spin of electrons[1]. Recently, spin-dependent electronic transport properties have attracted great attention due to their use in spintronic devices[2–4]. An extensive revolution has been brought to the study of magnetism with the development of two-dimensional magnetic materials (2DMMs), as the size effect has significant influence on the spin arrangement and electronic bands[5]. Magnetic materials with layered crystal structure have attracted great attention because of the feasibility of obtaining their atomically thin flakes by exfoliation, which enables the study of intrinsic magnetism in 2D limit as well as heterostructures composed of magnetic layers. So far, very few 2D ferromagnets are reported to exhibit Curie temperature ($T_C$) above room temperature, with most of them crystallizing in hexagonal lattice or its derivative, such as CrTe[6], $Cr_2Te_3$[7], and $Fe_3GaTe_2$[8]. This motivates us to explore room temperature 2D ferromagnetism in other lattice types.

$Mn_{2-x}T_xSb$ is one promising material family for 2D magnetism, where T is transition element such as V[9], Cr[10–15], Fe[16,17], Co[18–22], Cu[23,24], or Zn[25–27]. These materials crystalize in a tetragonal layered structure (space group $P4/nmm$). The parent compound $Mn_2Sb$ has two crystallographically different Mn atoms, Mn(I) and Mn(II), each forming a magnetic sub-lattice with antiparallel magnetic moments alignment in the magnetically ordered state. Below 550K, $Mn_2Sb$ shows a ferrimagnetic order with antiparallel Mn(I) and Mn(II) spins aligned collinearly along the $c$-axis. At 240K, there is another magnetic transition accompanied by spin reorientations toward the $ab$ plane for both Mn(I) and Mn(II)[28]. Other transition metal elements can substitute Mn(II), which not only tunes magnetic phases but also modifies cell volume[25], electric transport[20,25], and magnetocaloric effect[29], implying couplings between magnetism and other physical properties.

In this work, we studied non-magnetic Zn substitution for Mn(II) in $Mn_2Sb$, $Mn_{2-x}Zn_xSb$. Earlier studies on light Zn substitution ($x < 0.46$) have discovered multiple magnetic phases[25]. We have extended the study to the complete substitution up to $x = 1$, i.e., MnZnSb. Our measurements on $Mn_{2-x}Zn_xSb$ reveal distinct properties separated by a critical Zn composition of $x = 0.6$. The magnetism couples strongly with electronic properties and lead to distinct transport properties. In particular, a metal-to-insulator transition-like feature that appears to be first order transition is observed in Zn-rich samples, which can be tuned by magnetic field. Those discoveries, together with the feasibility of obtaining 2D flakes and the high temperature magnetism, provide further insight in 2D magnetism for spintronic applications.

## 2. Experimental Methods

All single crystals of $Mn_{2-x}Zn_xSb$, except for $Mn_2Sb$, were grown by a self-flux method using Zn flux. More than two years' efforts were devoted to identifying appropriate flux and optimize the growth procedures for various compositions. For each composition, Mn powder (99.6%, Alpha Aesar), Zn powder (99.9%, Alpha Aesar), and Sb powder (99.5%, Beantown Chemical) were loaded in an alumina crucible with the ratio of Mn:Zn:Sb = $1 + x : 6 : 1$; the crucible is then sealed in a quartz tube evacuated below $10^{-1}$ Pa. The tube was heated to the maximum temperature over 30 – 40 hours, held at this temperature for 3 days, cooled down to the centrifuge temperature and followed by subsequent centrifuge to remove the excess flux. The maximum growth temperature as well as the centrifuge temperature varied with the desired sample compositions. For example, the maximum and centrifuge temperatures for stoichiometric MnZnSb were 800 °C and 600 °C, respectively. Whereas for the $x \approx 0.4$ sample, these were 900 °C and 700 °C, respectively. For $Mn_2Sb$, single crystals were grown by using Bi flux with a ratio Mn : Bi : Sb

= 2 : 5 : 1. The source materials were heated to 950°C and centrifuged at 400°C. Millimeter-size $Mn_{2-x}Zn_xSb$ single crystals with metallic luster were obtained. The compositions of the obtained crystals were checked by energy dispersive X-ray spectroscopy (EDS), and we used such EDS compositions in the following discussions. The crystal structures were determined by single crystal XRD in a Bruker Apex I diffractometer with Mo Kα radiation ($\lambda = 0.71073$ Å). Magnetization and resistivity measurements were performed using a Physical Properties Measurement System (PPMS). To avoid any possible sample degradation issue, the single crystals were preserved in inert atmosphere inside a glovebox, and all the data used in this work were collected on fresh samples.

## 3. Results and Discussions:

As shown in Fig. 1a, $Mn_2Sb$ crystalizes in a layered tetragonal structure with two inequivalent Mn sites Mn(I) and Mn(II) [11,27,30]. Earlier neutron scattering studies have revealed that magnetic Mn(I) and Mn(II) form long range magnetic order below 550 K, which is characterized by inequivalent magnetic moments for Mn(I) and Mn(II) aligned antiparallelly along the *c*-axis [28]. Further lowering temperature to 240 K, Mn(I) and Mn(II) moments rotate toward *ab*-plane, leading to a low temperature ferrimagnetic phase. Zn doping has been found to substitute the Mn(II) sublattice (Fig. 1a) in lightly doped compositions [27] and MnZnSb [29,31]. In our $Mn_{2-x}Zn_xSb$ ($0 \leqslant x \leqslant 1$), single crystal *x*-ray structure analysis (see supplemental material) shows that Zn substitution maintains the tetragonal lattice symmetry with a space group *P*4/*nmm* (Fig. 1a), causing expanded in-plane lattice and reduced *c*-axis (Fig. 1b). Owing to the layered crystal lattice structure, thin flakes for $Mn_{2-x}Zn_xSb$ are accessible by mechanical exfoliation, as shown in Fig. 1c.

Our previous work on Zn-rich samples ($x > 0.5$) has revealed a magnetic transition that might be attributed to a spin rotation from the out-of-plane to the in-plane direction, which is similar to the scenario in parent compound $Mn_2Sb$ [31]. A giant topological Hall effect has also been observed, which reaches maximum in the $x \sim 0.85$ sample and is likely originating from a non-trivial spin texture [31]. In this work, the study is extended to the entire composition range with Zn content $0 \leqslant x \leqslant 1$, from which we observed unusual electronic transport properties. As shown in Fig. 2, $Mn_{2-x}Zn_xSb$ exhibits distinct transport properties below ($x < 0.6$, Fig. 2a) and above ($x > 0.6$, Fig. 2b) a critical composition of $x = 0.6$. For undoped $Mn_2Sb$ ($x = 0$), as shown in Fig. 2a, the temperature dependence of resistivity displays a good metallic behavior from 1.8 to 400 K with a residual resistivity ratio (RRR) $\rho(300K)/\rho(1.8K)$ of 25. This result is reproducible in a few pieces of $Mn_2Sb$ single crystals we tested, with some variation of residual resistivity and the corresponding RRR (from 25 to 36). Because our $Mn_2Sb$ single crystals were synthesized using a Bi flux, even though the EDS experiments did not probe the presence of Bi, a trace amount of Bi may slightly affect transport properties and cause the observed variation in the RRR. The observed metallic transport is consistent with the calculated metallic electronic band structure [32] and has also been observed in earlier reports [33,34]. Substituting Mn with Zn up to $x < 0.6$, the overall metallic transport persists, except for the emergence of a resistivity jump as indicated by the black arrows as shown in Fig. 2a. Such resistivity jump displays relatively weak composition dependence, gradually reducing from 128 K for the $x = 0.25$ sample to 105 K when $x = 0.59$. A thermal hysteresis is also observed for this resistivity jump, as shown in the inset of Fig. 2a.

Light metal substitution for Mn(II) sites in $Mn_2Sb$, i.e., $Mn_{2-x}T_xSb$ ($T$ = metal elements, $x \leqslant 0.4$), has been intensively studied. For example, the resistivity jump observed in our single crystals is consistent with the previous studies on polycrystalline Zn-substituted samples up to $x = $

0.2[26], which has also been observed in other doping studies using different dopants such as $Mn_{2-x}Cr_xSb$[11], $Mn_{2-x}Co_xSb$[20], and $Mn_{2-x}Cu_xSb$[23] with $x \leq 0.4$. All those earlier studies probed the emergence of a resistivity anomaly immediately upon substitution. Compared with those light doping studies, we have extended to Zn-rich samples up to $x = 1$ in this work. Interestingly, the transport properties change drastically when increasing Zn content beyond $x = 0.6$. Such abrupt change is reproducible in multiple samples with the compositions in the vicinity of $x = 0.6$. For the Zn-rich samples above x = 0.6, as shown in Fig. 2b, except for MnZnSb, these samples show complicated temperature dependence for resistivity. From 2 to 400 K, the temperature dependence can be divided into three regions separated by two transition-like features as indicated by the blue arrows (Fig. 2b). For example, for the $x = 0.61$ sample whose Zn content is only slightly above the critical value of 0.6, its resistivity increases with decreasing temperature at high temperatures from 400 K to 370 K, then starts to drop upon cooling until reaching 230 K. Below 230 K, resistivity again increases with decreasing temperature. In other words, the $x = 0.61$ sample displays metallic transport behavior in an intermediate temperature region (230 – 370 K), but shows non-metallic behaviors at high (400 to 370 K) and low (below 230 K) temperatures. The complicated non-metallic to metallic and again to non-metallic temperature dependence for resistivity is observed in all Zn-rich samples ($0.6 < x < 1$), except for MnZnSb (i.e., $x = 1$) which shows only one transition near 300 K (Fig. 2b). A thermal hysteresis has also been observed for the lower temperature transition, as shown in the inset of Fig. 2b.

Because $Mn_{2-x}Zn_xSb$ compounds are magnetic, it is natural to examine their magnetism and explore the possible connections with the unusual electronic transport properties. Magnetism for pristine $Mn_2Sb$ [28] and lightly ($x < 0.3$) Zn-substituted samples [25] have already been studied: $Mn_2Sb$ displays a ferrimagnetic order with Mn(I) and Mn(II) carrying opposite magnetic moments along the out-of-plane direction below 550 K (denote as HT-FI, see Fig. 3b). Further cooling to 240K, the magnetic moments for both Mn(I) and Mn(II) sublattices rotate to the in-plane direction (*a*- or equivalent *b*-axis), forming a low temperature ferrimagnetic phase (LT-FI, Fig. 3b). For slightly substituted samples, in addition to these two magnetic phases, Zn substitution leads to another magnetic phase transition occurring at lower temperature, which is characterized by the flipping of Mn magnetic moments to form a "weak ferrimagnetic" (WFI) phase [25]. As shown in Fig. 3b, for the LT-FI phase, Mn(I) and Mn(II) magnetic sublattices carry opposite moments, but the magnetic moments are parallel within each Mn(I) and Mn(II) magnetic sublattice [28]. However, in the WFI phase [25], moments within each Mn(I) and Mn(II) sublattice are not necessarily parallel. The entire magnetic structure can be understood in terms of the stacking of the Mn(II)-Mn(I)-Mn(II) tri-layer blocks along the *c*-axis. Within each block, Mn(I) and Mn(II) moments are antiparallel. Between blocks, the moment orientations alternate for every two blocks, as indicated by red and blue dashed boxes in Fig. 3b, which effectively form an antiferromagnetic structure.

Magnetism for light Zn-substitution ($x < 0.3$) samples have been studied using polycrystalline samples as described above [24,25]. The successful growth of $Mn_{2-x}Zn_xSb$ single crystals with Zn content $0 \leqslant x \leqslant 1$ allows us to build a complete magnetic phase diagram in the full composition range. Figure 3a summarizes the magnetic transition temperatures from temperature dependent susceptibility measurements (see Supplementary Material). The transition temperatures for Zn-rich samples ($x > 0.6$, represented by open blue diamonds) taken from our previous work [31] are also included for the purpose of showing the entire composition range. Systematic evolutions of HT-FI ordering temperature and spin reorientation temperature (i.e., HT-FI to LT-FI phase transition temperature) with Zn substitution are observed. For the WFI phase, which appears upon Zn substitution, interestingly, the transition temperature (orange triangles, Fig. 3a) does not display a strong composition dependence but abruptly disappears beyond $x = 0.6$. In Fig. 3a, we also depict the transition temperatures extracted from resistivity measurements (Fig. 2). In Zn-less samples ($x < 0.6$), though multiple magnetic transitions have been established by magnetic susceptibility measurements, these magnetic phase transitions are not necessarily manifested in clear features in resistivity measurements. For example, a resistivity jump occurs coincidently with the formation of the WFI phase, but resistivity is featureless during the spin-reorientation (HT-FI to LT-FI) transitions. For Zn-rich samples ($x > 0.6$), resistivity transitions match well with the magnetic transitions, as shown in Fig. 3a.

The absence of the resistivity feature during spin-reorientation transition in lightly substituted samples is widely observed in other transition metal substitution studies such as $Mn_{2-x}Cr_xSb$ [11], $Mn_{2-x}Co_xSb$ [20], and $Mn_{2-x}Cu_xSb$ [23,24] but the mechanism is unknown. One possible explanation is that the spin reorientation changes the moment orientations but leaves the magnetic wavevector unaffected. Therefore, the electronic states may not be disturbed sufficiently to result in a clearly observable feature in transport. In contrast, at the WFI transition, the formation of magnetic lattices changes the periodicity (Fig. 3b), which might lead to a super zone gap [23] that causes the resistivity jump. In addition, the thermal hysteresis at the WFI transition in resistivity measurements (Fig. 2a, inset) implies the first-order nature of the transition. Indeed, a first-order structure transition accompanied with WFI transition has been observed for a low Zn composition ($x = 0.1$) [25]. Such structure transition preserves the tetragonal symmetry and is characterized by reduced $c$-axis and expanded $a$ and $b$-axes [25]. A structure transition is naturally expected to affect transport properties. Owing to the difficulty in synthesizing single crystals for samples with very low Zn content, we were unable to determine the critical Zn composition to induce such WFI phase. The previous studies based on polycrystalline samples found that such WFI phase can be easily induced by small amount of dopants such as Zn, Cu, Co, and Cr [11,20,23–26]. The effect of substitution on modifying transport properties is also manifested in our magnetotransport measurements. As shown in Figs. 4 and 5, the pristine $Mn_2Sb$ displays a positive magnetoresistance (MR) which is expected for normal metal without strong spin scattering. In addition, it is quite interesting that MR increases with rising temperature (Fig. 4). The MR for $Mn_2Sb$ does not show typical quadratic field dependence for orbital MR, but rather a sublinear behavior with a broad dip at low field up to 395K, the highest temperature we measured (Fig. 5a). Such low field dip is reminiscent weak antilocalization, but it is not expected to occur at such a high temperature. Therefore, the unusual increase of MR at high temperature as well as the low field dip, might be associated with the magnetism of the $Mn_2Sb$, which strongly implies a coupling between magnetism and transport. Because $Mn_2Sb$ undergoes magnetic transition at 550 K [28], measurements at higher temperatures covering that magnetic transition is needed to better understand the unusual MR. For the Zn-substituted samples, distinct from the pristine $Mn_2Sb$, they all display clear negative MR for the entire temperature range, implying enhanced spin scattering that can be attributed to the magnetic fluctuations due to the disturbance of long-range magnetic order by Zn substitution. Furthermore, because the structure transition for the WFI phase reduces $c$-axis [25], this consequently enhances the magnetic coupling of the adjacent Mn(I) and Mn(II) layers. Therefore, lattice and spin degrees of freedom are strongly coupled in this Zn-less $Mn_{2-x}Zn_xSb$, which further affects transport properties. Such coupling of charge transport, spin, and lattice implies an efficient approach to engineering properties. For example, our magnetotransport measurements have revealed a shift of the WFI transition to lower temperatures upon applying magnetic field. An example for the $x$ = 0.30 sample is shown in Fig. 4. Such suppression of transition temperature further leads to large negative magnetoresistance below the WFI transition temperature in lightly Zn-substituted samples, as shown in Fig. 5.

The observed weak composition dependence for WFI phase transition temperature is quite different from magnetic doping studies such as $Mn_{2-x}Fe_xSb$ [17] and $Mn_{2-x}Co_xSb$[20,21], which might be attributed to the non-magnetic dopant Zn. WFI phase disappears beyond $x = 0.6$, which might be ascribed to the fact that the magnetic interaction between Mn(I) and Mn(II) is drastically modulated when Mn(II) is substantially substituted by non-magnetic Zn. As discussed above, WFI phase appears to be characterized by enhanced Mn(I) – Mn(II) magnetic interactions, so that it might be unfavorable when Mn(II) magnetism is suppressed. Having said that, why the WFI phase disappears so abruptly at $x = 0.6$ is unclear and needs more theoretical and experimental (such as neutron scattering) efforts.

As WFI phase disappears for heavy Zn substitution ($x > 0.6$), consequently, there are two magnetic phase transitions for those Zn-rich samples as shown in the phase diagram (Fig. 3). Our previous magnetization study suggests that these two magnetic phases in Zn-rich samples should still be ferrimagnetic [31]. Unlike the lightly Zn-substituted samples, the magnetic structures for Zn-rich samples (except for MnZnSb) have not been experimentally determined. Therefore, we label the high temperature and low temperature ferrimagnetic phases for Zn-rich samples as HT-FI' and LT-FI', respectively, as shown in Fig. 3a. Previous neutron scattering studies on $Mn_2Sb$ [11], lightly substituted samples [27], and completed substituted MnZnSb [35] have revealed a ferromagnetic Mn(I) magnetic sublattice with out-of-plane moments at high temperatures, which undergoes a spin reorientation toward in-plane direction at lower temperature. Therefore, though magnetic structure for Zn-rich samples ($0.6 < x < 1$) has not been reported, it is reasonable to assume similar magnetic orders and transitions for Mn(I) sublattice. For the Mn(II) sublattice, which is heavily substituted by non-magnetic Zn, the remaining Mn(II) are still expected to be magnetically coupled with Mn(I). Previous band structure calculations have revealed that the major contribution to the density of state near Fermi level DOS($E_F$) in $Mn_2Sb$ [36] and MnZnSb [37] is mainly from Mn(I) *d*-orbitals, so Zn substitution to Mn(II) mainly modifies magnetism. Accompanied by the disappearance of WFI phase in those Zn-rich samples is the drastic change of transport properties. As mentioned above, the non-metallic transport behavior (Fig. 2b) in the high temperature paramagnetic state deviates strongly from the metallic electronic band structure [37]. Since the crystal lattice does not change with temperature and upon Zn substitution (see Supplementary Material), the non-metallic transport might be associated with strong spin scatterings. Mn(II) magnetism is heavily disturbed by non-magnetic Zn substitution, so its coupling to Mn(I) may cause strong spin fluctuations that incoherently scatter conductions electrons. Such a scenario might be further fostered by the fact that $Mn_{2-x}Zn_xSb$ is a layered material with DOS($E_F$) mainly from Mn(I). The negative MR (Fig. 5) also indicates the existence of spin scattering in Zn-substitute samples. When long range magnetic order is formed in the HT-FI' phase, spin scattering is expected to be suppressed. As shown in Fig. 2b, metallicity is enhanced in the HT-FI' phase in Zn-rich samples, as reflected by the reduced negative slope d$\rho$/d$T$, or even the positive slope which is expected for metals.

In addition, non-metallic transport in magnetic metals have also been observed in other materials such as $Fe_{1+y}Te$. $Fe_{1+y}Te$ possesses a layered structure with Fe square lattice layers, displaying a bad metal behavior above the AFM ordering temperature and metallic transport in the AFM state [38]. Magnetic fluctuations can be induced by doping additional Fe to the interstitial site near the Fe plane, which enhances carrier localization [38][39]. The bad metallic behavior in $Fe_{1+y}Te$ has been attributed to spin freezing, i.e., Hund's couplings tend to align spins of the electrons in different orbitals and enhance local moment, which couples with itinerant electrons in a manner similar to Kondo effect [40,41]. Despite the similarities in transport and magnetic properties in our Zn-rich $Mn_{2-x}Zn_xSb$ and $Fe_{1+y}Te$, whether a similar spin freezing mechanism applies needs further theoretical and experimental investigations.

In the low temperature LT-FI' phase, the Zn-rich $Mn_{2-x}Zn_xSb$ becomes non-metallic again. A sharp resistivity jump occurs at the LT-FI' phase transition temperature (Fig. 2b), which is accompanied by a thermal hysteresis (Fig. 2b, inset). Such resistivity jump is not observed in MnZnSb ($x = 1$) in which Zn fully replaced the Mn(II) sublattice and the sample only displays a room temperature ferromagnetic order owing to the ordering of the Mn(I) sublattice. Though our single crystal XRD experiments at various temperatures (see Supplementary Material) do not reveal a change of lattice symmetry for the LT-FI' phase, those Zn-rich samples might still undergo a structure transition that maintains the tetragonal symmetry in a mannar similar to the WFI transition in Zn-less samples mentioned above. In addition, the LT-FI' phase transition can also be suppressed by applying magnetic field as shown in Fig. 4c, which is also similar to the WFI phase in Zn-less samples (Fig. 4b). Despite of these similarities, the resistivity jumps in Zn-rich ($0.6 < x < 1$) samples are much sharper and resembles a metal-to-insulator transition. Such metal-to-insulator transition-like behavior might be due to the cell multiplication for magnetic cells in the LT-FI' phase, or caused by the formation of superstructure due to the ordering in the Mn(II)-Zn mixed plane. Having said that, it is worth noting that the resistivity of the samples is still low ($0.1 \sim 1$ mΩ cm) after the resistivity jump, so the samples might not enter an truly insulating state. Whether such sharp resistivity jump is caused by a possible gap or by a sudden change in scattering mechanism needs further experimental and theoretical efforts.

## 4. Conclusion

In conclusion, we have successfully obtained single crystals for $Mn_{2-x}Zn_xSb$ ($0 \leqslant x \leqslant 1$) and studied their electronic and magnetic properties. We found that $Mn_{2-x}Zn_xSb$ possesses a critical composition of $x = 0.6$ and displays rich but distinct properties for Zn-less and Zn-rich samples. Below the critical Zn content, $Mn_{2-x}Zn_xSb$ is metallic and characterized by three magnetic phases, but only the low temperature WFI phase exhibit features in transport properties. Above the critical Zn content, the WFI phase disappears abruptly and the transport properties modify drastically. A metal-to-insulator transition-like accompanied by the formation of a low temperature magnetic order behavior is observed, which can be tuned by applying magnetic field. The transition is first order, implying a possible structure transition. Our observations indicate strong coupling between spin, charge, and lattice in this material. Owing to the feasibility of obtaining 2D flakes for $Mn_{2-x}Zn_xSb$, our study provides a brand-new material platform that integrates high temperature (above room temperature) magnetism and exotic transport phenomena, which are tunable by composition and magnetic field.

## 5. Acknowledgement

Work at the University of Arkansas was supported by the MonArk NSF Quantum Foundry supported by the National Science Foundation Q-AMASE-i program under NSF award No. DMR-1906383. Z.Q.M. acknowledges the support from NSF under Grant No. DMR 2211327. H.C. acknowledges support from NSF award No. DMR-1848281. J.H. acknowledges the support from NSF under Grant No. DMR-2238254.

**Figures**

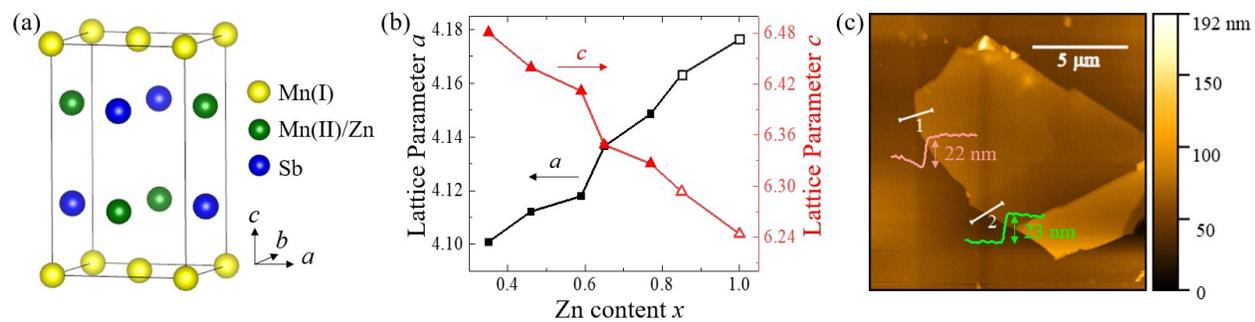

FIG. 1. (a) Crystal Structure for $Mn_{2-x}Zn_xSb$. (b) Composition dependence of lattice parameter *a* and *c* for $Mn_{2-x}Zn_xSb$ obtained from *x*-ray diffraction. (c) atomic force microscope topography image of an exfoliated MnZnSb flake. Insets: line scans showing the flake thickness.

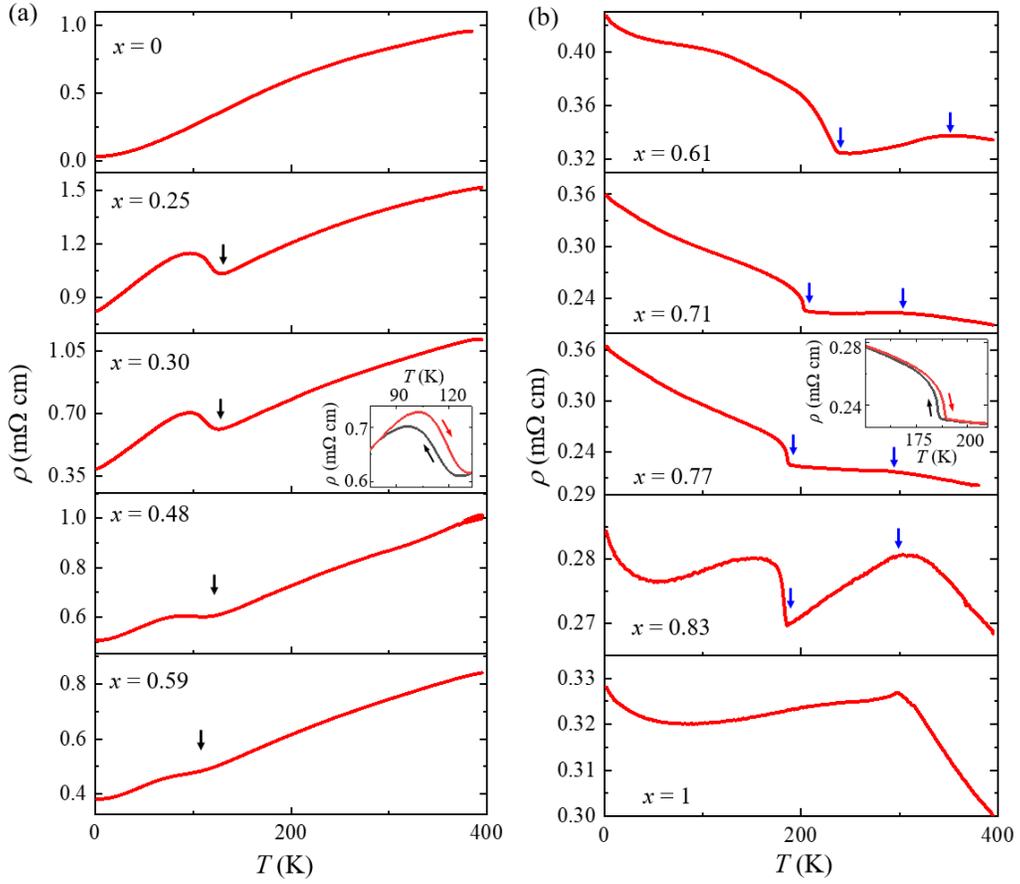

FIG. 2. Temperature dependence of resistivity for (a) Zn-less ($x < 0.6$) and (b) Zn-rich ($x > 0.6$) $Mn_{2-x}Zn_xSb$ samples. The inset shows thermal hysteresis.

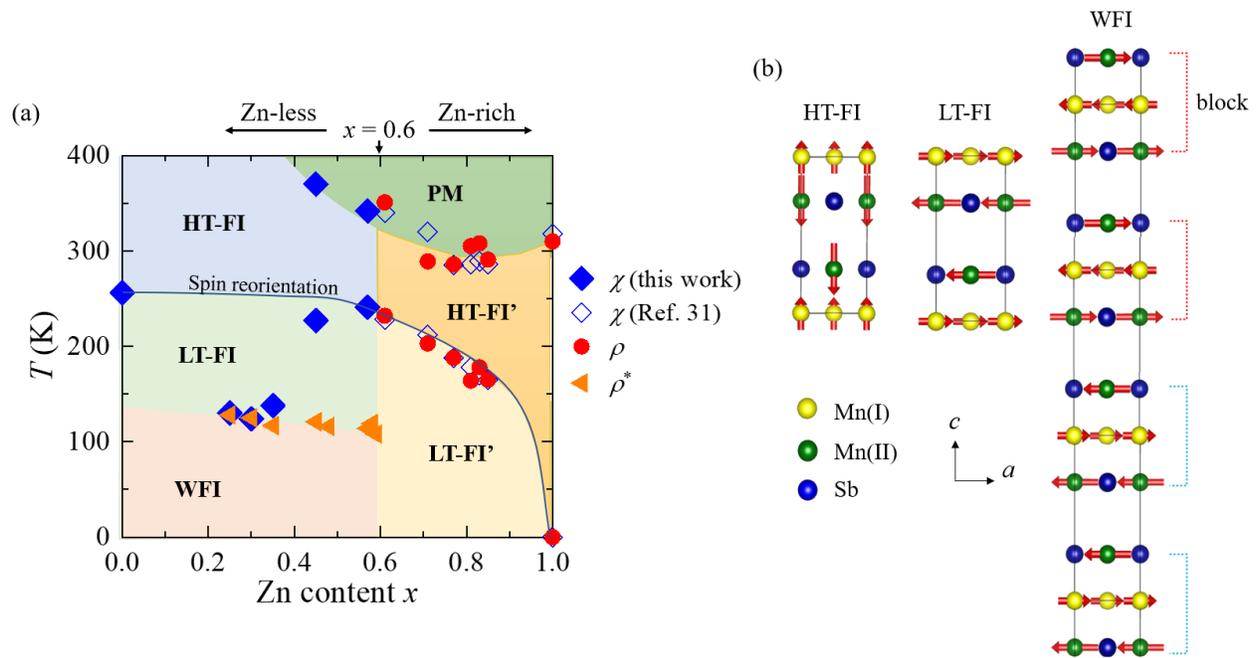

FIG. 3. (a) Magnetic phase diagram for $Mn_{2-x}Zn_xSb$ constructed from magnetic susceptibility $\chi(T)$ (see Supplementary Material) and resistivity $\rho(T)$ data. (b) The schematic magnetic structures for HI-FI, LT-FI, and WFI phases.

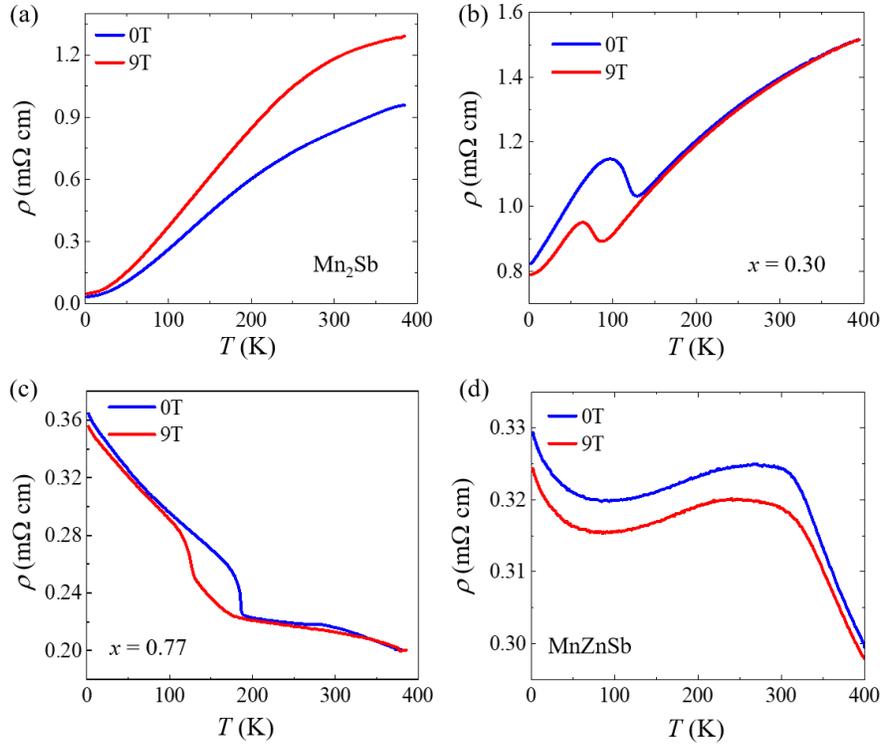

FIG. 4. Temperature dependence of resistivity measured at 0 and 9 T magnetic field for (a) $Mn_2Sb$, (b) $x = 0.30$, (c) $x = 0.77$, and (d) MnZnSb samples.

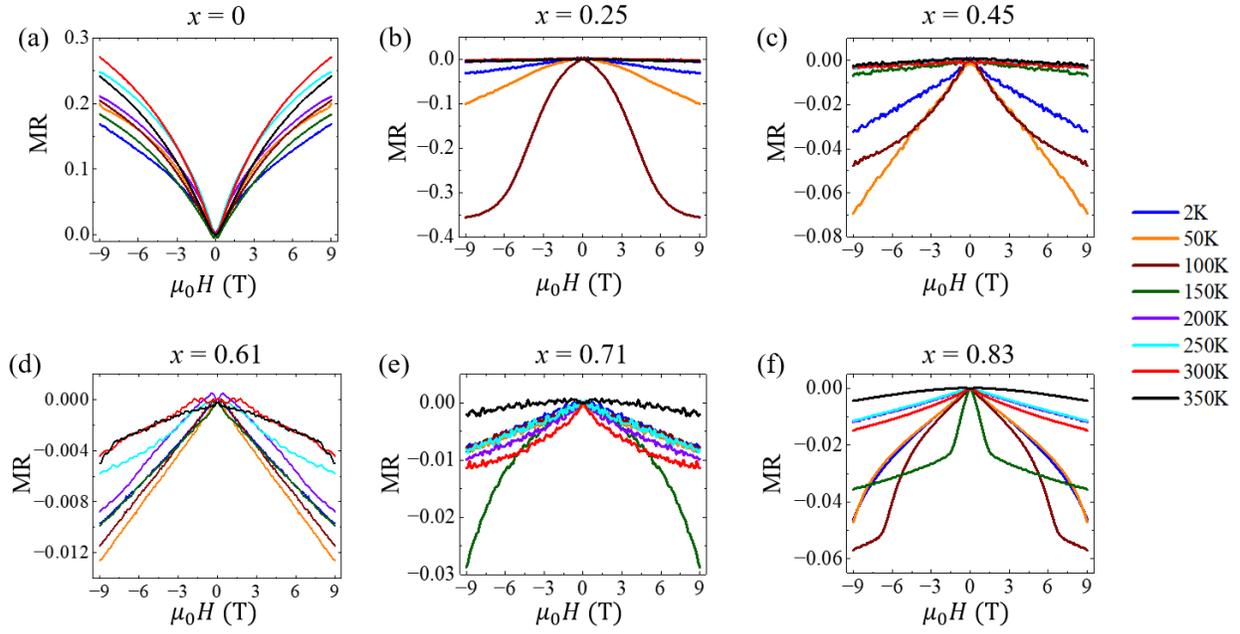

FIG. 5. Normalized magnetoresistivity, MR = $[\rho_{xx}(H) - \rho_{xx}(H=0)]/\rho_{xx}(H=0)$ for (a) $x = 0$, (b) $x = 0.25$, (c) $x = 0.45$, (d) $x = 0.61$, (e) $x = 0.71$, and (f) $x = 0.83$ samples.